\documentclass[12pt,a4paper,dvips]{article}
\usepackage{epsfig,wrapfig,times,mathptm} 
\renewcommand{\section}[1]{\vspace{6pt} \noindent\mbox{#1} \newline \noindent}
\renewcommand{\subsection}[1]{\vspace{6pt} \noindent\mbox{\underline{#1}} 
\newline \noindent}
\renewcommand{\subsubsection}[1]{\vspace{6pt} \noindent\mbox{\underline{#1}}
\noindent}

\newfont{\sans}{cmss10}

\begin{document}
{\small OG 2.2.13 \vspace{-24pt}\\}     
{\center \LARGE SEARCH FOR TeV COUNTERPARTS IN GAMMA RAY BURSTS 
\vspace{6pt}\\}
P.J.Boyle$^1$,
J.H.Buckley$^2$,
A.M. Burdett$^3$,
J.Buss\'{o}ns Gordo$^1$,
D.A.Carter-Lewis$^4$,
M.Catanese$^4$,
M.F.Cawley$^5$,
V.Connaughton$^6$,
D.J.Fegan$^1$,
J.P.Finley$^7$,
J.A.Gaidos$^7$,
K.Harris$^2$,
A.M.Hillas$^3$,
R.C.Lamb$^8$,
F.Krennrich$^4$,
R.W.Lessard$^7$,
C.Masterson$^1$,
J.E.McEnery$^1$,
G.Mohanty$^4$,
N.A.Porter$^1$,
J.Quinn$^{1,2}$,
A.J.Rodgers$^3$,
H.J.Rose$^3$,
F.W.Samuelson$^4$,
G.H.Sembroski$^7$,
R.Srinivasan$^7$,
T.C.Weekes$^2$,
J.Zweerink$^4$,
\vspace{6pt}\\
{\it $^1$ Physics Department, University College, Dublin, Ireland\\
$^2$Whipple Observatory, Harvard-Smithsonian CfA, USA\\
$^3$ Physics Department, University of Leeds, United Kingdom\\
$^4$ Department of Physics and Astronomy, Iowa State University, USA\\
$^5$ Physics Department, St.Patrick's College, Maynooth, Ireland\\
$^6$ Marshall Space Flight Center, Alabama, USA\\
$^7$ Department of Physics, Purdue University, Indiana, USA\\
$^8$ Space Radiation Laboratory, Caltech, Pasadena, USA\\
\vspace{-12pt}\\}

{\center ABSTRACT\\}

Based on BACODINE network notification the Whipple Observatory
gamma-ray telescope has been used to search for the delayed TeV
counterpart to BATSE-detected gamma-ray bursts. In the fast slew mode,
any point in the sky can be reached within two minutes of the burst
notification. The search strategy, necessary because of the
uncertainty in burst position and limited FOV of the camera, is
described.

\section{INTRODUCTION}

The Whipple collaboration has an active observational program which is
dedicated to searching for TeV counterparts of classical gamma ray
bursts.  Since May 1994, the Whipple Observatory 10 meter gamma-ray
telescope (Cawley, et al., 1990) has made 16 follow-up observations of
BATSE burst notifications, but to date no positive identifications
have been made (Connaughton, et al., 1995, Connaughton, et al.,
1997a). The field of view of our current instrument is limited to 3.5
degrees and the error box for the BATSE position of gamma-ray bursts
is 10 degrees in diameter. Therefore, more than a single pointing has
to be used for the TeV follow up after a gamma-ray burst. The current
observation technique is to take a 28 minute exposure at the initial
location and 4 more exposures 3 degrees away, with a final exposure
taken again at the initial position as in Figure 1a (Connaughton, et
al., 1997a).  In order to find any associated TeV component a two
dimensional (2D) search strategy has been developed in order to locate
the precise region of emission (Akerlof, et al., 1993, Connaughton, et
al., 1997a, Lessard, et al., 1997).

Assuming gamma ray bursts have an associated TeV component, it is
uncertain as to just how long after the BATSE notification of a
gamma-ray burst that this emission would last. At present, we cover
less than $50\%$ of the actual BATSE error box in 2.5 hours and of the
16 follow-ups observed to date, only 6 had full observations of all
positions (Table 1). All other follow-ups were terminated after an
hour due to sun/moonrise or the position being too low to track. Owing
to the uncertainty in the timescale of bursts and the lack of coverage
of the whole error box, a non-detection cannot be used as an argument
against the possibility of gamma-ray burst emission extending up to
TeV-emission energies.

\vspace{12pt}

{\footnotesize
\begin{table}[h]
\vspace{-12pt}
\begin{center}
\caption{Whipple Observations of BACODINE Positions}
\begin{minipage}{6in}
\begin{tabular}{cccccccc}
\hline
\hline
Date & BATSE & BACODINE & Difference & Durat. & Whipple & Elev- & Cycles \\
& Trig.\footnote{Trigger number of BATSE burst}
 & Intensity & BACO- & BATSE & delay & ation & observed \\
& & Cts/s\footnote{Number of counts/s measured by BATSE in the
first 1 or 2 seconds of the burst.}
 & Hunts. (deg)\footnote{Angular difference between the burst
coordinates provided through BACODINE and the final estimated
burst position from Huntsville.}
 & (s)\footnote{Duration of the burst seen by
BATSE.} 
 & (min)\footnote{Length of time that elapsed between
sending the BACODINE message and the start of the first Whipple
run.} 
 & (deg)\footnote{Average elevations of the first 
and last positions covered by the 10-metre telescope.}
 & (1-6)\footnote{Whipple coverage of burst area.
The numbers listed indicate the positions covered in the observations.
The 6 positions are described in the text and are shown
in Figure 1a.  An entry of "complete" means all 6
positions and their control observations were completed.}  \\
\hline
940516 & 2980 & 630 & N/A & N/A & 19 & 24-14 & 1 (1 hour) \\ 
950208 & 3408 & 778 & 1.64 & N/A & 16 & N/A & 1,2  \\
950405 & 3494 & 704 & 9.5 & N/A & 8 & 27-33 & 1,2,3+cont \\
950524 & 3598 & 8726 & 1.37 & 6 & 5 & 56-31 & complete \\
950625 & 3649 & 1661 & 3.84 & 40 & 18 & 28-41 & 1-6 \\
950701 & 3658 & 9134 & 2.3 & 15 &  56 & 41-69 & complete \\
951117 & 3909 & 1955 & 10.1 & 25 & 5 & 30-24 & 1 (1 hour) \\
951119 & 3911 & 801 & 6.98 & 60 & 20 & 45-54 & complete \\
951124 & 3918 & 1231 & 5.98 & 150 & 2 & 59-76 & complete \\
951220 & 4048 & 6698 & 2.88 & 17 & 27 & 71-37 & complete \\
960521 & 5467 & 2919 & 9.13& & 14 & 80-68 & 1,2 \\ 
961017 & 5634 & 3193 & 2.74 & 41 & 7 & 28-23 & complete \\
961111\footnote{Trigger \#5665 was a terrestial trigger} & 5665 & 496 & - & -  & 4 & 54-80 & 1-4 \\
961206 & 5706 & 1312 & 6.4 & 1.4 & 5 & 26-22 & 1-3 \\
970304 & 6113 & 8842 & 3.1 & 11 & 3 & 28-42 & 1-4 \\
970308 & 6117 & 1120 & 13.6 & 2 & 4 & 43-72 & complete \\
\hline
\end{tabular}
\end{minipage}
\end{center}
\end{table}}

\section{RASTER SEARCH TECHNIQUE}

In order to achieve a better coverage of the 10 degree error box we
now scan a circular region around the BATSE burst position with a
constant angular velocity. A raster scan across parallel lines in
declination (Figure 1b) with a spacing of 1 degree results in complete
coverage of the BATSE error box. A continuous motor speed of 0.1
degree/s facilitates a scan of the whole error box within 400 seconds.
The fast scanning method has the advantage that we may give a flux
upper limit within the first 300-400 seconds or less depending on the
final improved BATSE burst position. Although the sensitivity within
the first 300 seconds is not great, a burst of the calibre of 910503
or 930131 would be easily detected if the spectrum extends from the
GeV region up to TeV energies ( Hurley 1996). The time scale of those
two bursts was 84 and 100 seconds respectively and only if the
burst positions are close to the actual telescope position can a
resonable coverage of the BATSE error box be achieved.

Recent improvements have lowered the time for response to a BACODINE
notification of the Whipple telescope to within 15 seconds of the
BATSE trigger. A new tracking system for the telescope will be
implemented in the fall of 1997, making the most distant point on the
sky accessible within 2 minutes.

\begin{figure}
\centerline{\hbox{
\epsfig{file=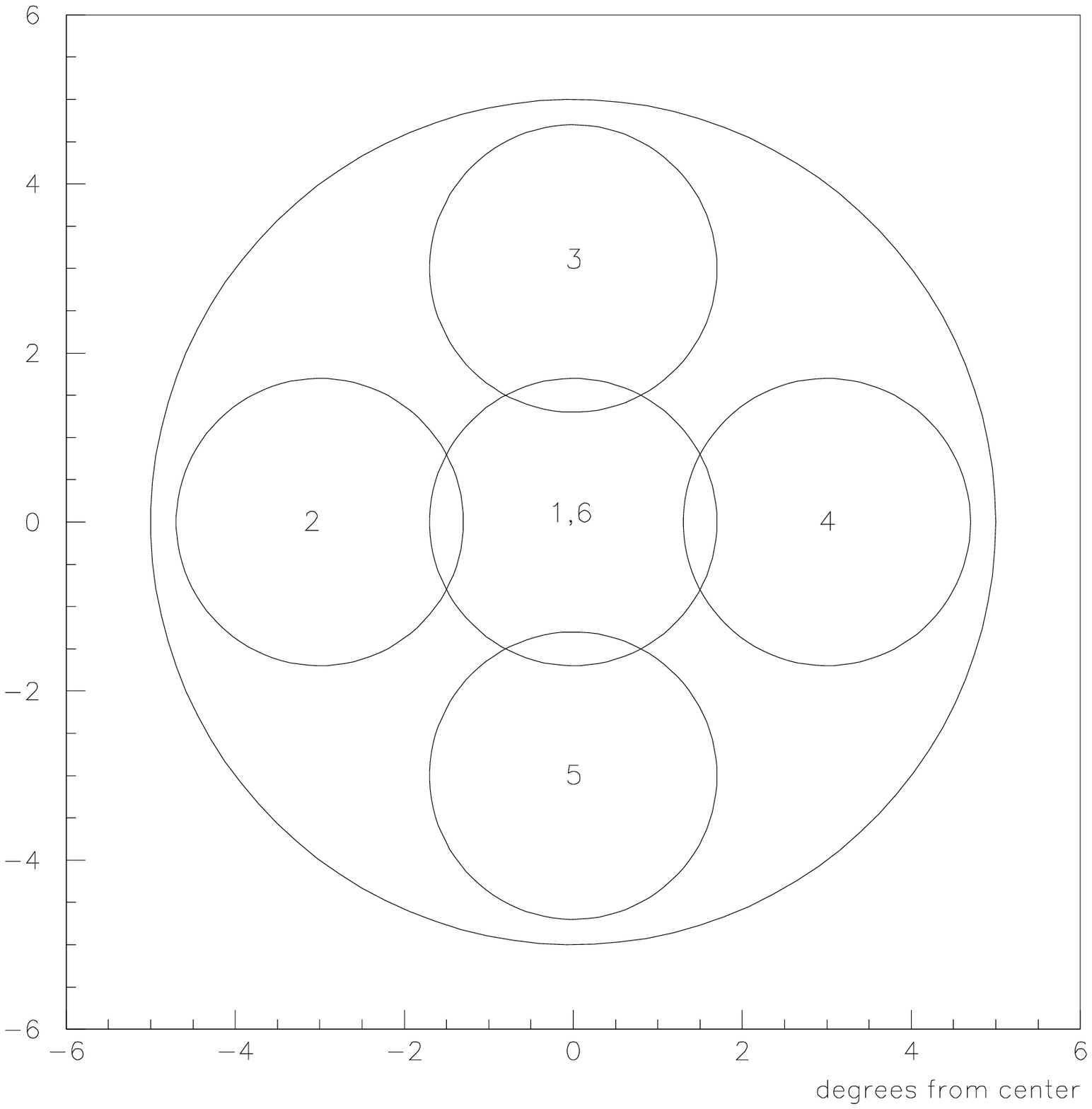,width=7.5cm}
\epsfig{file=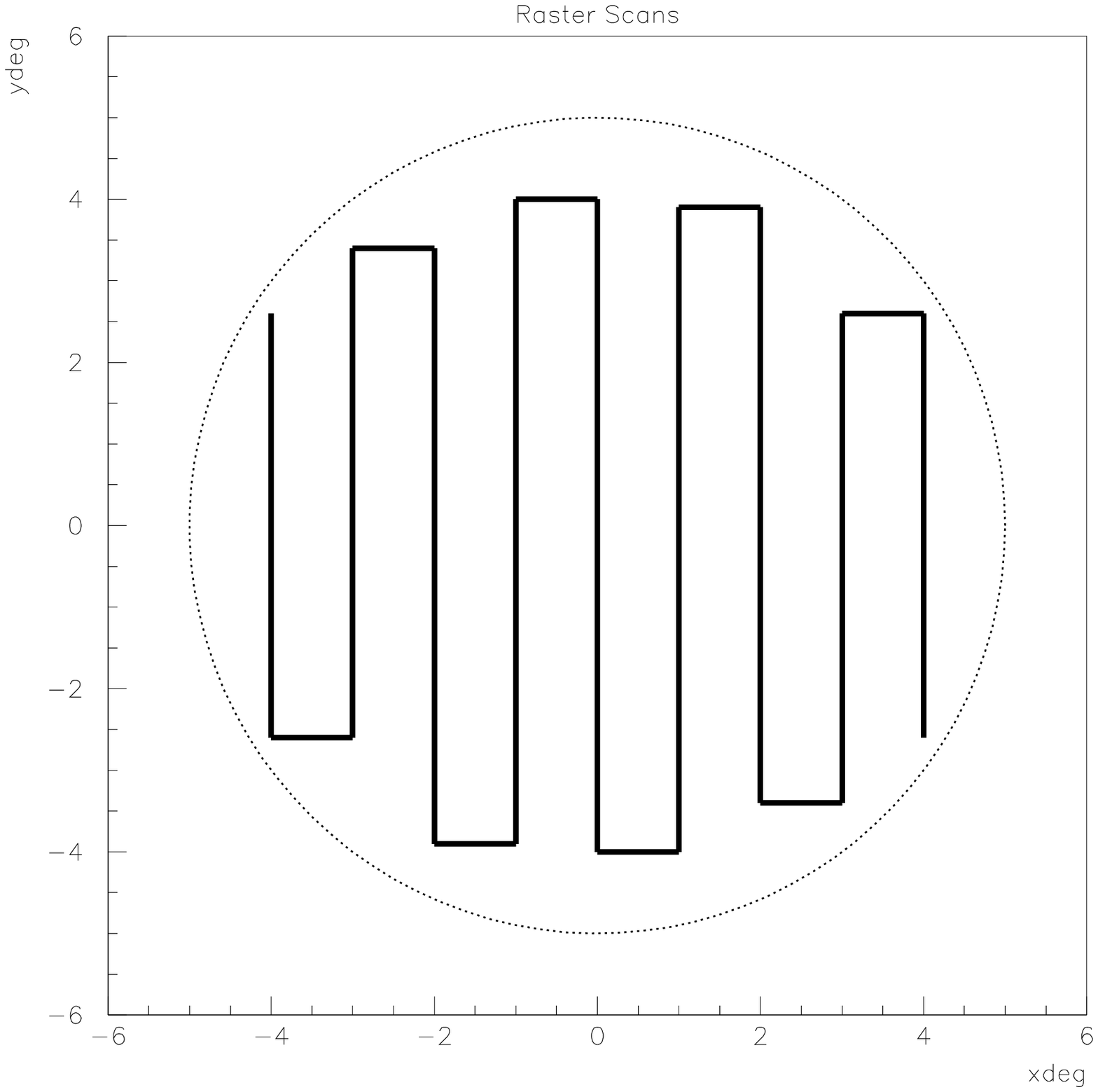,width=7.5cm}}}
\caption{\it (a): Current Response to BACODINE notifications. Each
circle represents the field of view of the 10-meter reflector at the 5
different positions following a BATSE burst. (b): Raster Scan
response. The reflector is moved at a constant velocity continuously
along the track to cover the BATSE error box.}

\end{figure}

A burst seen by BATSE and EGRET on 940217 lasted for some 1.5 hours at
GeV energies. This class of gamma-ray burst might easily have been
missed by the previous method but it could now be observed with a
reasonable sensitivity (9 minutes effective observation time assuming
a 3 degree FOV and a 10 degree error box).

The result of each raster scan will be analysed and displayed by a
quicklook analysis to search for a strong excess. If there is no
significant excess this scanning process can be continued for 2-5
hours if possible. If something significant is apparent, the
telescope will be pointed at this position.

\begin{wrapfigure}[19]{r}{7.5cm}
\epsfig{file=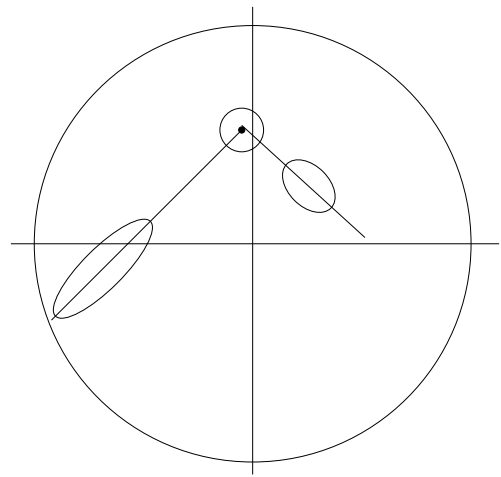,width=6.5cm}
\caption{\it 2-D reconstruction of points of origin of
showers. For small impact parameters, the image has a structure
close to that of a circle, whereas with increasing impact parameter it 
appears more elongated. }
\end{wrapfigure}

\subsection{ANALYSIS}

The Whipple high energy gamma-ray telescope images \v{C}erenkov light
from air showers onto an array of 151 photomultipliers covering a 3.5
degree field of view. The imaging technique uses information of the
angular spread and orientation to reject more than 99.7\% of the
cosmic-ray induced showers while retaining over 50\% of the possible
gamma-ray induced events.  For discrete objects the source is placed
at the center of the field of view and candidate gamma-ray events
(CGRE) are selected on the basis of the a combination of (a) image
shape ({\it length} \& {\it width}) and (b) orientation of the major
axis of the shower (Reynolds, et al., 1993). However, as a burst
source has an unknown position a 2-D search can determine the incident
direction of the CGRE. The search makes use of the orienation,
elongation and asymmetry of the image.  Monte Carlo studies have shown
that gamma-ray images are a) aligned towards the source position b)
elongated in proportion to their impact parameter on the ground and c)
asymetrical, images are weighted towards their point of origin (Figure
2). The distance between the image centroid and the true point of
origin of a shower is represented by the imaging parameter {\it disp}:

\begin{equation}
\label{poo}
 disp  = 1.85 -1.85(width/length)
\end{equation}

from which a unique point of origin can be calculated. As each event
represents in itself a unique telescope position in Right Ascension
(RA) and Declination (Dec), each point of origin must be translated
into an absolute co-ordinate grid. Each point on the grid is then
tested for an excess of gamma-ray-like events. The off center
properties of the Whipple gamma-ray telescope are well established
(Connaughton, et al., 1997b) and are used to determine a flux at any
point in the camera.

\section{ACKNOWLEDGEMENTS} We acknowledge the technical support of
Emmet Roache. This research is supported by grants from the US
Department of Energy and by NASA, by PPARC in the UK, and by Forbairt
in Ireland.

\section{REFERENCES}
\setlength{\parindent}{-5mm}
\begin{list}{}{\topsep 0pt \partopsep 0pt \itemsep 0pt \leftmargin 5mm
\parsep 0pt \itemindent -5mm}
\vspace{-15pt}
\item Akerlof, C.W., et al., 1991, Ap.J. 377, L97
\item Cawley, M.F., et al., 1990, Exp. Astron., 1, 185.
\item Connaughton, V., et al., 1995, Proc. 24th ICRC, 2, 96.
\item Connaughton, V., et al., 1997a, Ap.J., 479, 859.
\item Connaughton, V., et al., 1997b, Astroparticle Physics (submitted).
\item Hurley, K., Space Science Reviews, v.75, p.43-52.
\item Lessard, R.W., et al., 1997, Towards a Major Atmos. Cerenkov Detector - V.
\item Reynolds, P.T., et al., 1993, Ap.J. Letters 404, 206

\end{list}

\end{document}